\documentclass[12pt,aps,prb,preprint,showpacs,showkeys]{revtex4}  

\usepackage{amsmath}    
\usepackage{amsfonts}   
\usepackage{amssymb}
\usepackage{graphicx}   
\usepackage{subfigure}

\begin{document}
\baselineskip = 7.5mm \topsep=1mm
\begin{center}{\LARGE\bf Monte Carlo simulation of the three-dimensional XY model with bilinear-biquadratic exchange interaction}
\vspace{5 mm}\\ H.Nagata$^{\rm a}$, M.\v{Z}ukovi\v{c}$^{\rm b}$ and T.Idogaki$^{\rm a*}$ \vspace{5 mm}\\
\end{center}
$^{\rm a}$ Department of Applied Quantum Physics, Kyushu University
\newline
$^{\rm b}$ Institute of Environmental Systems, Kyushu University \vspace{3 mm}
\newline
\noindent {\bf{Abstract.}} The three-dimensional XY model with bilinear-biquadratic exchange
interactions $J$ and $J'$, respectively, has been studied by Monte Carlo simulations. From the detailed
analysis of the thermal variation of various physical quantities, as well as the order parameter and
energy histogram analysis, the phase diagram including two different ordered phases has been determined.
There is a single phase boundary from a paramagnetic to a dipole-quadrupole ordered phase, which is of
second order in a high $J/J'$ ratio region, changing to a first-order one for $0.35 \leq J/J' \leq 0.5$.
Below $J/J'=0.35$ there are two separate transitions: the first one to the quadrupole long-range order
(QLRO) phase at higher temperatures, followed by another one to the dipole-quadrupole long-range order
(DLRO) phase at lower temperatures. The finite-size scaling analysis yields values of the critical
exponents for both the DLRO and QLRO transitions close to the values for the conventional XY model which
includes no biquadratic exchange. \vspace{20 mm} \\ $PACS\ codes$: 75.10.Hk; 75.30.Kz; 75.40.Cx;
75.40.Mg.
\newline
$Keywords$: XY model; Bilinear-biquadratic exchange; Phase transition; Quadrupole ordering; Histogram
Monte Carlo simulation;\\ \vspace{20 mm} \\ $*$Corresponding author.
\newline
Permanent address: Department of Applied Quantum Physics, Faculty of Engineering, Kyushu University,
Fukuoka 812-8581, Japan
\\ Tel.: +81-92-642-3810; Fax: +81-92-633-6958
\newpage
\noindent {\bf\Large{1.Introduction}} \vspace{3mm}
\newline
\indent The problem of biquadratic (or generally higher-order) interactions has attracted much attention
for several decades now. For systems with Heisenberg symmetry and spin $S=1$ it has been tackled by mean
field approximation (MFA) \cite{chen-levy1}, high-temperature series expansion (HTSE) calculations
\cite{chen-levy2}, as well as within a framework of some other approximative schemes
\cite{micnas,chaddha99}. The case of $S>1$ has also been treated by MFA \cite{sivar72,sivar73}. Those
studies have shown that the biquadratic interactions can induce various interesting properties such as
tricritical and triple points, quadrupole ordering, separate dipole and quadrupole phase transitions
etc. The problem of the biquadratic interactions in systems with XY spin symmetry, however, has has
received much less attention. The case $S=1$ has been addressed, however, only in a high $J/J'$ ratio
region where the biquadratic exchange has no significant influence on phase transitions
\cite{chaddha96}. Chen $et.al$ for the first time looked into the problem of the critical exponents for
the phase transitions in the classical XY model with the bilinear-biquadratic exchange. They used HTSE
to calculate transition temperatures and critical exponents for cubic lattices in the region of $J/J'
\geq 1$ \cite{chen-etal1}. However, based on the MFA assumption that for $J/J'<1$ the transition to the
dipole long-range order (DLRO) phase is of first order, they limited their calculations in this region
only to the separate quadrupole long-range order (QLRO) transitions taking place for $0<J/J \leq 0.35$
\cite{chen-etal2}. Here we note, however, that the rigorous proof of the existence of dipole and
quadrupole long-range order at finite temperature on the classical bilinear-biquadratic exchange model
has been provided only recently \cite{tanaka,campbell}.
\newline
\indent In the present paper we focus on the region of comparatively low $J/J'$ $(\leq1)$, which is the
most interesting from the point of view of the critical behaviour but, at the same time, the least
elucidated. We use standard Monte Carlo (SMC) and histogram Monte Carlo (HMC) simulations, and
investigate the possible kinds of long-range ordering, their nature, and critical exponents, for a
classical XY ferromagnet with biquadratic exchange on a simple cubic lattice. The obtained phase diagram
captures all important features induced by the biquadratic exchange such as separate dipole and
quadrupole ordering, first-order transitions, and consequently the triple and tricritical points
appearance. Furthermore, we perform a finite-size scaling (FSS) analysis in order to calculate the
susceptibility and correlation length critical exponents, $\nu$ and $\gamma$, respectively, for both
DLRO and QLRO transitions.\vspace{6mm}
\newline
\noindent {\bf\Large{2.Model and Monte Carlo simulation}}\vspace{3mm}
\newline
\indent The classical XY model with bilinear-biquadratic exchange interactions can be described by the
Hamiltonian
\begin{equation}
H=-J\sum_{\langle i,j \rangle}\mbox{\boldmath $S$}_{i} \cdot \mbox{\boldmath $S$}_{j} -J'\sum_{\langle
i,j \rangle}(\mbox{\boldmath $S$}_{i} \cdot \mbox{\boldmath $S$}_{j})^{2} \ ,
\end{equation}
where $\mbox{\boldmath $S$}_{i} = (S_{ix},S_{iy})$ is a two-dimensional unit vector at the $i$th lattice
site , $\langle i,j \rangle$ denotes the sum over nearest neighbors, and $J,\ J'>0$ are the bilinear and
biquadratic exchange interaction constants, respectively. It is known that such a spin system displays
long-range ordering of both dipole and quadrupole moments. The order parameters corresponding to the
respective kinds of ordering are the dipole long-range order (DLRO) and the quadrupole long-range order
(QLRO) parameters, $m$ and $q$, respectively, defined by

\begin{equation}
\label{eq.m}m=\langle M \rangle/N,\ {\mathrm{where}}\
M=\left[\left(\sum_{i}S_{ix}\right)^{2}+\left(\sum_{i}S_{iy}\right)^{2}\right]^{\frac{1}{2}}\ ,
\end{equation}
\begin{equation}
\label{eq.q}q=\langle Q \rangle/N,\ {\mathrm{where}}\
Q=\left[\left(\sum_{i}\left(\left(S_{ix}\right)^{2}-\left(S_{iy}\right)^{2}\right)\right)^{2}+
\left(\sum_{i}2S_{ix}S_{iy}\right)^{2}\right]^{\frac{1}{2}}\ ,
\end{equation}
where $N$ is the total number of the lattice sites, and $\langle \cdots \rangle$ denotes the thermal
average. The respective orders are schematically depicted in Fig.1. DLRO corresponds to the
ferromagnetic directional arrangement of spins while QLRO represents an axially ordered state in which
spins can point either direction along the axis of ordering. Obviously, DLRO always includes QLRO and,
hence, DLRO actually represents dipole-quadrupole long-range order.
\newline
\indent In our simulations we first perform standard Monte Carlo (SMC) simulations on systems of the
linear lattice size up to $L$ = 24, assuming periodic boundary condition throughout. Spin updating
follows a Metropolis dynamics and averages are calculated using $1\times10^{4}$ Monte Carlo steps per
spin (MCS/s) after equilibrating over another $5\times10^{3}$ MCS/s. Besides DLRO and QLRO parameters
$m$ and $q$, we calculate the system internal energy $E$, and the specific heat per site $c$, calculated
from energy fluctuations by
\begin{equation}
\label{eq.c}c=\frac{(\langle E^{2} \rangle - \langle E \rangle^{2})}{Nk_{B}T^{2}}\ ,
\end{equation}
the susceptibility per site $\chi_{O}$, calculated from LRO parameters fluctuations by
\begin{equation}
\label{eq.chi}\chi_{O} = \frac{(\langle O^{2} \rangle - \langle O \rangle^{2})}{Nk_{B}T}\ ,
\end{equation}
and the fourth-order long-range order (LRO) cumulant (Binder parameter) $g_{O}$ as
\begin{equation}
\label{eq.g} g_{O}=2-\frac{\langle O^{4}\rangle}{\langle O^{2}\rangle^{2}}\ ,
\end{equation}
where $O$ stands for the respective parameters $M$ and $Q$.
\newline
\indent Temperature dependence of these quantities gives us an estimate of the location, as well as
nature of a transition. First-order transitions are usually accompanied by discontinuities in order
parameters and energy, and hysteresis when cooling and heating. If transition is second order, it can be
roughly located by the $c$ peak position or, alternatively, by the position of the fourth-order LRO
cumulant curves intersection for various lattice sizes.
\newline
\indent In order to obtain more reliable and more precise data, we further perform histogram Monte Carlo
(HMC) calculations, developed by Ferrenberg and Swendsen \cite{ferr-swen1,ferr-swen2}, at the transition
temperatures estimated from the SMC calculations for each lattice size. Here we also treat larger
lattice sizes (up to $L$ = 30), and thermal averages are taken over $2\times10^{6}$ MCS after discarding
another $1\times10^{6}$ MCS used for bringing the system into the thermal equilibrium. We calculate the
energy histogram $P(E)$, the order parameters histograms $P(O)$\ $(O=M,Q)$, as well as the logarithmic
derivatives of $O$ and $O^{2}$ with respect to $K=1/k_{B}T$, which can be written in the form
\begin{equation}
\label{eq.D1}D_{1O} = \frac{\partial}{\partial K}\ln\langle O \rangle = \frac{\langle OE
\rangle}{\langle O \rangle}- \langle E \rangle\ ,
\end{equation}
\begin{equation}
\label{eq.D2}D_{2O} = \frac{\partial}{\partial K}\ln\langle O^{2} \rangle = \frac{\langle O^{2} E
\rangle}{\langle O^{2} \rangle}- \langle E \rangle\ .
\end{equation}

\noindent Further, we use the histograms in order to determine FSS behaviour which allows us to extract
the critical exponents. In the case of a second-order transition, the extrema of the calculated
thermodynamic quantities are known to scale with a lattice size as:
\begin{equation}
\label{eq.scalchi}\chi_{O,max}(L) \propto L^{\gamma_{O}/\nu_{O}}\ ,
\end{equation}
\begin{equation}
\label{eq.scalV1}D_{1O,max}(L) \propto L^{1/\nu_{O}}\ ,
\end{equation}
\begin{equation}
\label{eq.scalV2}D_{2O,max}(L) \propto L^{1/\nu_{O}}\ ,
\end{equation}

\noindent where $\nu_{O}$ and $\gamma_{O}$ represent the correlation length and susceptibility critical
exponents, respectively. In the case of a first-order transition (except for the order parameters), they
display a volume-dependent scaling, $\propto L^{3}$. \vspace{6mm}
\newline
\noindent {\bf\Large{3.Phase boundaries and transition order}}\vspace{3mm}
\newline
\indent The temperature dependences of the specific heat and the DLRO parameter in the region where the
biquadratic exchange is less or equal to the bilinear one, namely $J/J'$ = 10, 2.5, and 1.0, are shown
in Fig.2. Observing the specific heat peaks we can see that with decreasing exchange ratio $J/J'$ the
transition temperature $T/J$ is raised. In this region both dipole and quadrupole moments order at the
same temperature and, therefore, there is only one phase transition from the disordered paramagnetic
phase to the DLRO phase. This state of a single phase transition persists also for lower exchange ratio
values down to $J/J' \simeq 0.35$. Below $J/J' \simeq 0.35$ quadrupoles start ordering separately at
temperatures higher than those for dipole ordering. Thus the phase boundary branches and a new middle
phase of axial quadrupole long-range order (QLRO) without magnetic dipole ordering opens between the
paramagnetic and DLRO phases, and it broadens as $J/J'$ decreases. In Figs.3(a,b,c) we present the
temperature variation of the specific heat, the DLRO and QLRO parameters $m$ and $q$, respectively, and
the corresponding susceptibilities $\chi_{M}$ and $\chi_{Q}$ at $J/J'=0.2$. We can see that here
quadrupoles order before dipoles, forming a fairly broad region of QLRO without DLRO. The snapshots of
the spin states in the respective phases appearing as temperature is lowered are depicted in Fig.3d.
\newline
\indent Further our concern will be the question of what order these transitions are. As mentioned
earlier, a first-order transition is manifested by discontinuous behaviour of the order parameter and
energy and, hence, the two quantities should display a bimodal (double-peak) distribution at the
transition. On the other hand, if a transition is second order, only a single-peak distribution is
observed. One should make sure, however, that the lattice size is sufficiently large and the single-peak
behaviour does not result from finite size effects. We calculate the energy and order parameter
histograms at the critical temperatures previously estimated by the SMC calculations for various lattice
sizes. For $0.5 < J/J'$ only a singe-peak distribution is found, suggesting a second-order transition,
in agreement with continuously looking temperature variation of both the order parameter and energy in
this region. If the exchange ratio is lowered, the double-peak structure of the energy and order
parameter histograms appears. Using the Lee and Kosterlitz method \cite{lee-kosterlitz} we can adjust
temperature to make the two peaks equally high and, such a way, precisely determine the transition
temperature for a given lattice size (Fig.4). Fig.5 shows the energy distribution diagrams for $J/J'$ =
0.35, 0.4 and 0.5, and various lattice sizes with the respective size-dependent transition temperatures
$T_{c}(L)$. As can be seen from Figs.5(a,c), i.e. the cases of a comparatively weak first-order
transitions near multicritical points, the bimodal distribution can only be observed at sufficiently
large $L$. On the other hand, in the case of $J/J'=0.4$ (Fig.5(b)), the dip between the peaks is
observable already at smaller $L$, quite rapidly approaching zero as $L$ is increased, indicating
discontinuous behaviour of the energy at a rather strong first-order transition. If a transition is
first order, $T_{c}(L)$ should scale with volume as
\begin{equation}
\label{eq.scalT}|T_{c}-T_{c}(L)| \propto L^{-d} \ ,
\end{equation}
where $d$ is the system dimension. In Fig.6 we plot $T_{c}(L)$ vs $L^{-3}$, using the scaling relation
(\ref{eq.scalT}), and the values of $T_{c}(L)$ extrapolated to $L\longrightarrow \infty$ give us fairly
precise estimates of the real transition temperatures for the respective $J/J'$, as follows:
$T_{c}/J'=1.2018(1)$ for $J/J'=0.35$, 1.2918(3) for 0.4, and 1.4851(2) for 0.5. In the region of the
separate QLRO and DLRO transitions ($0 < J/J' < 0.35$) we found no double-peak energy distribution for
neither kind of transition. However, there are noticeable differences in thermodynamic quantities
behaviour between the QLRO and DLRO transition. While in the case of the QLRO transition the energy and
QLRO parameter show apparent continuous behaviour even at fairly large lattice sizes, in the case of the
DLRO transition, although we could not observe any discontinuities nor hysteresis, the observed slopes
are extremely sharp (Fig.3(b)), which is also reflected to the spike-like specific heat and
susceptibility peaks (Figs.3(a,c)). This tendency is even more pronounced as the lattice size is
increased. Therefore, we cannot exclude possibility of a discontinuous behaviour and, hence, a
first-order transition, for $L\longrightarrow \infty$. Another way to decide the order of the transition
is by analyzing the temperature dependence of the Binder parameter $g_{O}(L,T)$. In the case of a
first-order transition it should display a minimum after entering a paramagnetic phase \cite{vollmayr}.
In our case, the DLRO parameter seemingly displays such a behaviour, in contrast to the QLRO one
(Fig.7), however the transition is not to a disordered paramagnetic but another ordered - the QLRO
phase. Moreover, the minima do not scale with volume as is should be at a first-order transition and,
therefore, they should not be seen a sign of a first-order transition. Unfortunately, unlike for the
case of the QLRO transition, which has already been predicted to be of second order for a
three-dimensional XY model \cite{carmesin}, there are no previous theories on what order the DLRO
transition should be in this region. The resulting phase diagram for the region of $0 \leq J/J' \leq 1$
is drawn in Fig.8. For the sake of comparison we also included the HTSE calculations results
\cite{chen-etal1,chen-etal2}. We can see that in spite of the relatively small lattice sizes used in our
calculations the critical temperature values match quite well those obtained from the HTSE calculations.
However, the DLRO transition temperature values within $0 \leq J/J' < 1$ were not previously calculated
and, hence, here, our data present completely new results. \vspace{6mm}
\newline
\noindent {\bf\Large{4.Finite-size scaling analysis}}\vspace{3mm}
\newline
\noindent {\it{4.1. Critical exponents at DLRO transition}} \\ \indent Besides phase boundaries, we also
investigated how the biquadratic exchange can modify the critical exponents at a second-order
transition. We first perform a finite-size scaling for the DLRO transitions with $J/J' = \infty$, 2.5, 1
and 0.8, and calculate the correlation length and susceptibility critical exponents $\nu_{M}$ and
$\gamma_{M}$, respectively, associated with DLRO. We note that the case of $J/J' = \infty$, i.e. when
the biquadratic exchange is absent, has previously been calculated by MC simulations \cite{janke}, but
we included it also in our calculations just for the sake of comparison. The results for $J/J' = 0.8$
are presented in Fig.9 in ln-ln plot. The slopes yield values of $1/\nu_{M}$ for the logarithmic
derivatives $D_{1M},\ D_{2M}$ and $\gamma_{M}/\nu_{M}$ for the susceptibility $\chi_{M}$. Next, we plot
the size-dependent transition temperatures $T_{c}(L)$, determined from the peak positions of various
quantities, versus $L^{-1/\nu_{M}}$. The value of $\nu_{M}$ is taken as an average of the two values
obtained from the logarithmic derivatives. The data then should fit straight lines, which extrapolated
to $L\longrightarrow \infty$ should converge to a single point - the real $T_{c}$ (Fig.10). The obtained
values of $T_{c}$, $\gamma_{M}/\nu_{M}$ and $1/\nu_{M}$ are listed in Table 1, comparing with those
previously calculated in Ref. \cite{janke} for $J/J' = \infty$ and by the HTSE method \cite{chen-etal1}.
As seen from the table, the exponents are modified by the presence of the biquadratic exchange in
continuous manner, which is in agreement with the HTSE calculations.\vspace{3mm}
\newline
\noindent {\it{4.2. Critical exponents at QLRO transition}} \\ \indent In the region of the QLRO
transition, we examined the critical exponents for transitions at $J/J' =$ 0, 0.1, 0.2 and 0.3. Plots
similar to those in Fig.9 are drawn in Fig.11 for the case of $J/J' =$ 0.1, this time in order to obtain
the slopes corresponding to the values of $1/\nu_{Q}$ for the logarithmic derivatives $D_{1Q},\ D_{2Q}$
and $\gamma_{Q}/\nu_{Q}$ for the susceptibility $\chi_{Q}$. From the obtained values of $\nu_{Q}$ we
plot the size dependence of the transition temperature (Fig.12).  Finally, the extrapolated values of
$T_{q}(L\longrightarrow \infty)$ are calculated for the respective values of $J/J'$. In Table 2 the
resulting values are listed and compared to those from the HTSE calculation \cite{chen-etal2}.
Reasonable agreement between the present results and the HTSE results is achieved in both the critical
exponents and transition temperatures cases. \vspace{3mm}
\newline
\noindent {\it{4.3. Critical exponents in multicritical point vicinity}}\\ \indent At $J/J' \simeq 0.35$
the frontiers of the paramagnetic, DLRO and QLRO phases merge into a single point. The vicinity of this
point presents a crossover region between first- and second-order transitions which should be reflected
to the critical exponents' behaviour. In Fig.13 we present the results for $J/J'$ = 0.35. Here, the
values of all slopes are significantly enhanced compared to the values for a second-order transition,
and for the case of $J/J'$ = 0.38 (Fig.14) quite close to the limiting value 3, which should be reached
at a first-order transition. Observing the histograms issued at $J/J'$ = 0.35 (Fig.5(a)) and 0.38, one
would conclude, however, that the transitions are of a first order. The fact that the slopes are less
than 3 should not be considered as a discrepancy, since the calculated slopes only present the effective
values affected by finite size effects. The data in Figs.13 and 14 appear to lie on curves turning
upwards, indicating that within the present sizes a true linear regime has not yet been established and
still larger sizes would be needed to bring the system into such a regime. The calculated slopes are
summarized in Table 3.\vspace{6mm}
\newline
\noindent {\bf\Large{5.Concluding remarks}}\vspace{3mm}
\newline
\indent We studied effects of the biquadratic exchange on the phase diagram of the classical XY
ferromagnet on a simple cubic lattice. We tried to cover all significant critical phenomena induced by
the presence of the biquadratic exchange and bring a solid picture of a role of this higher-order
exchange interaction in the critical behaviour of the considered system. In the region where the
bilinear exchange is dominant we found only one phase transition to the DLRO phase, which remains second
order until the exchange ratio reaches the value $J/J' \simeq 0.5$. Upon further lowering of the ratio,
the transition changes to a first-order one at the tricritical point and remains this way down to $J/J'
\simeq 0.35$. Below this value the phase boundary splits into the QLRO transition line at higher
temperatures and the DLRO transition line at lower temperatures. While the QLRO transition is clearly of
second order, the order of the DLRO transition, due to ambiguous behaviour of the physical observables
at the transition, could not be established with certainty, although, a second-order transition seems to
be more likeable. We consider performing some more simulations on different lattices of larger sizes and
longer simulation time in order to obtain reliable data for the scaling analysis, which should
eventually provide a conclusive answer to the question of the order of the DLRO transition in the
considered region. Finite-size scaling analysis showed that the critical exponents display a slight
variation with changing $J/J'$ but for neither DLRO nor QLRO transitions significantly deviate from the
standard three-dimensional XY universality class values. \vspace{6mm}
\newline
\noindent {\bf\Large{Acknowledgments}}\vspace{3mm}
\newline
\indent We wish to thank Dr. A. Tanaka for valuable discussions concerning the theoretical background of
the studied problem and Dr. Y. Muraoka for numerous technical consultations when running the simulations
on the supercomputer.

\newpage
%

\begin{figure}[!t]
\includegraphics[scale=0.5]{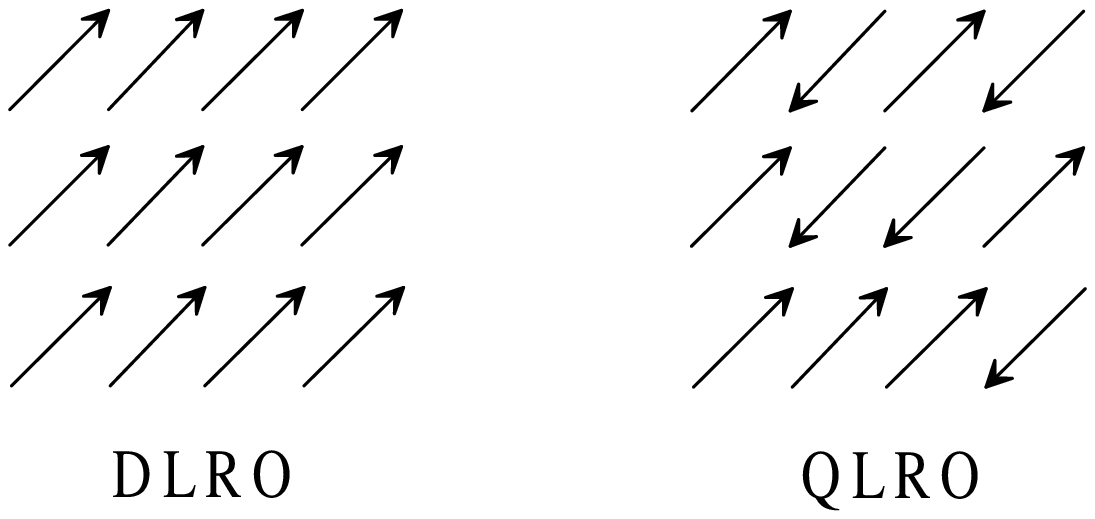}
\caption{Schematic picture of DLRO and QLRO states.}
\end{figure}

\begin{figure}[!t]
\subfigure{\includegraphics[scale=0.5]{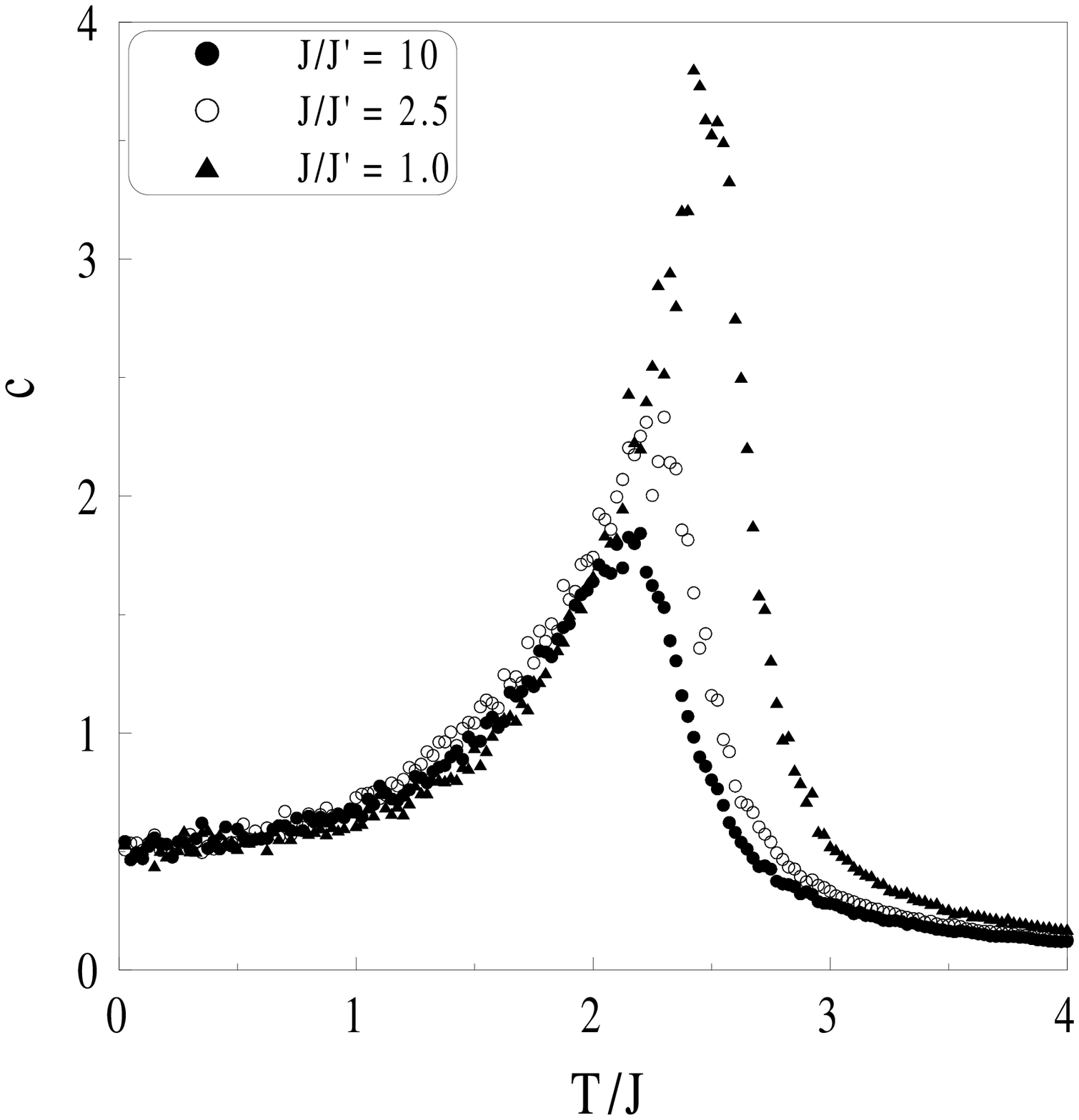}}
\subfigure{\includegraphics[scale=0.5]{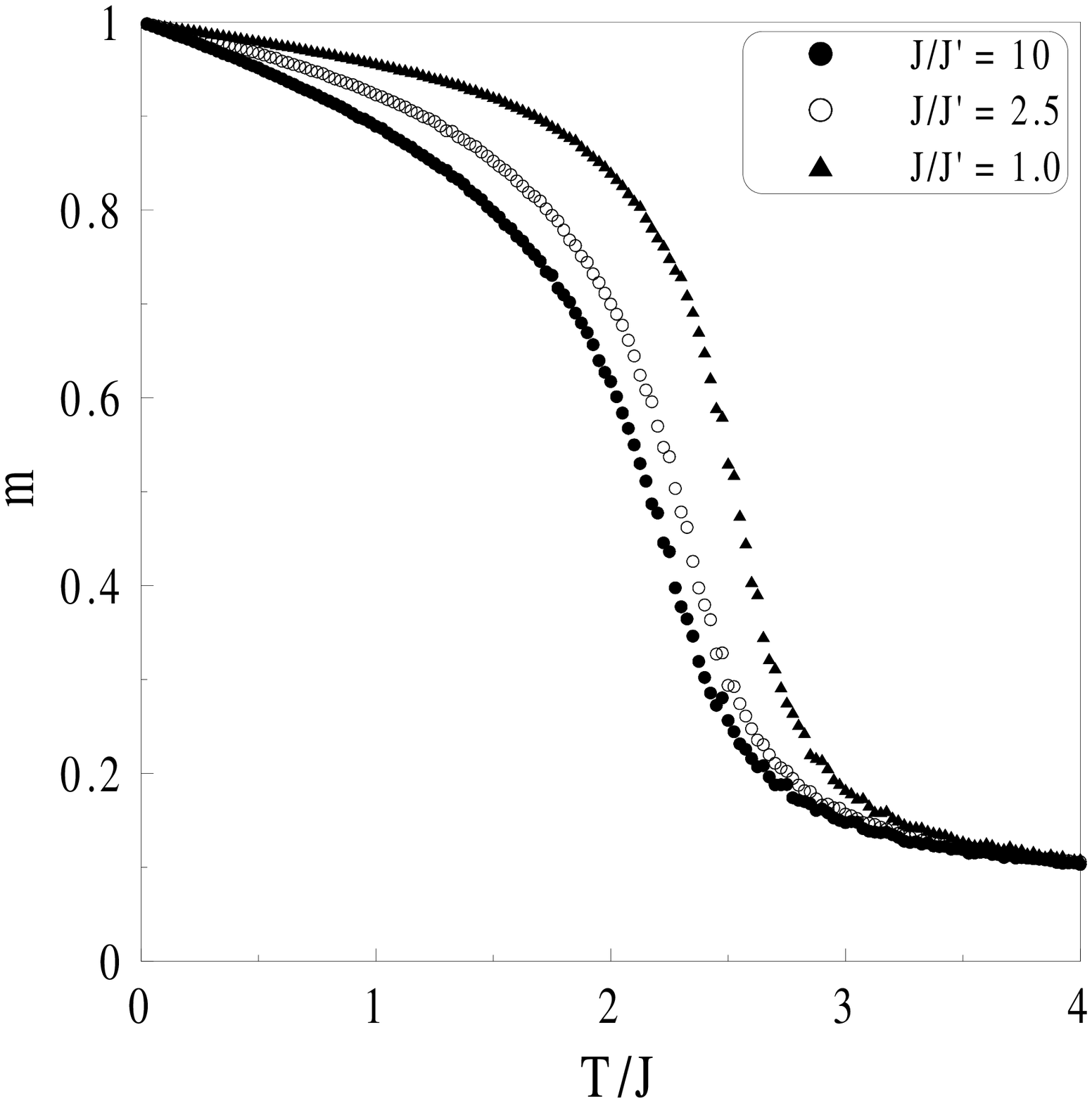}}
\caption{Temperature dependence of (a) the specific heat and (b) the order parameter for
$J/J'$ = 1.0, 2.5 and 10.}
\end{figure}

\begin{figure}[!t]
\subfigure{\includegraphics[scale=0.4]{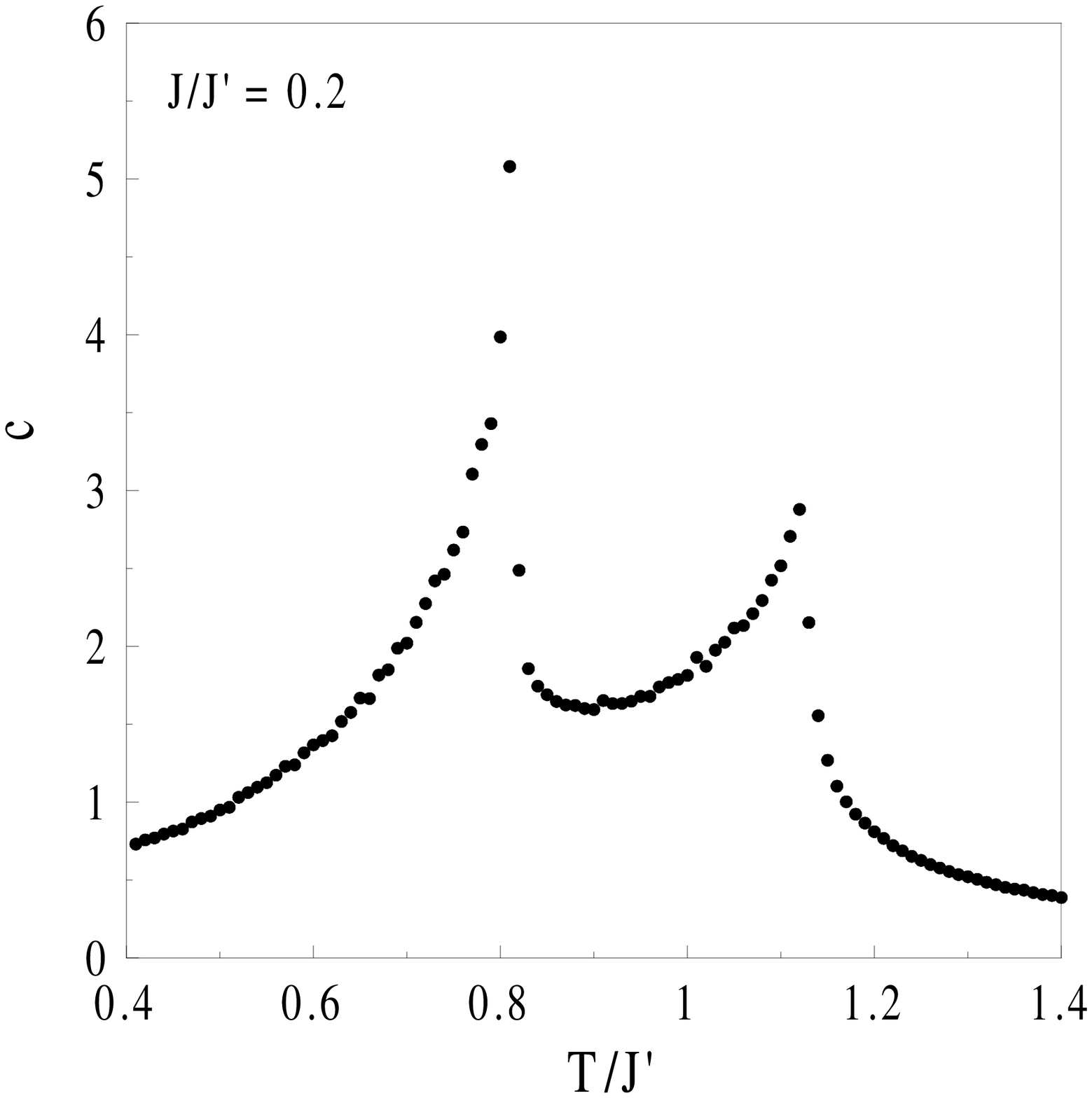}}
\subfigure{\includegraphics[scale=0.4]{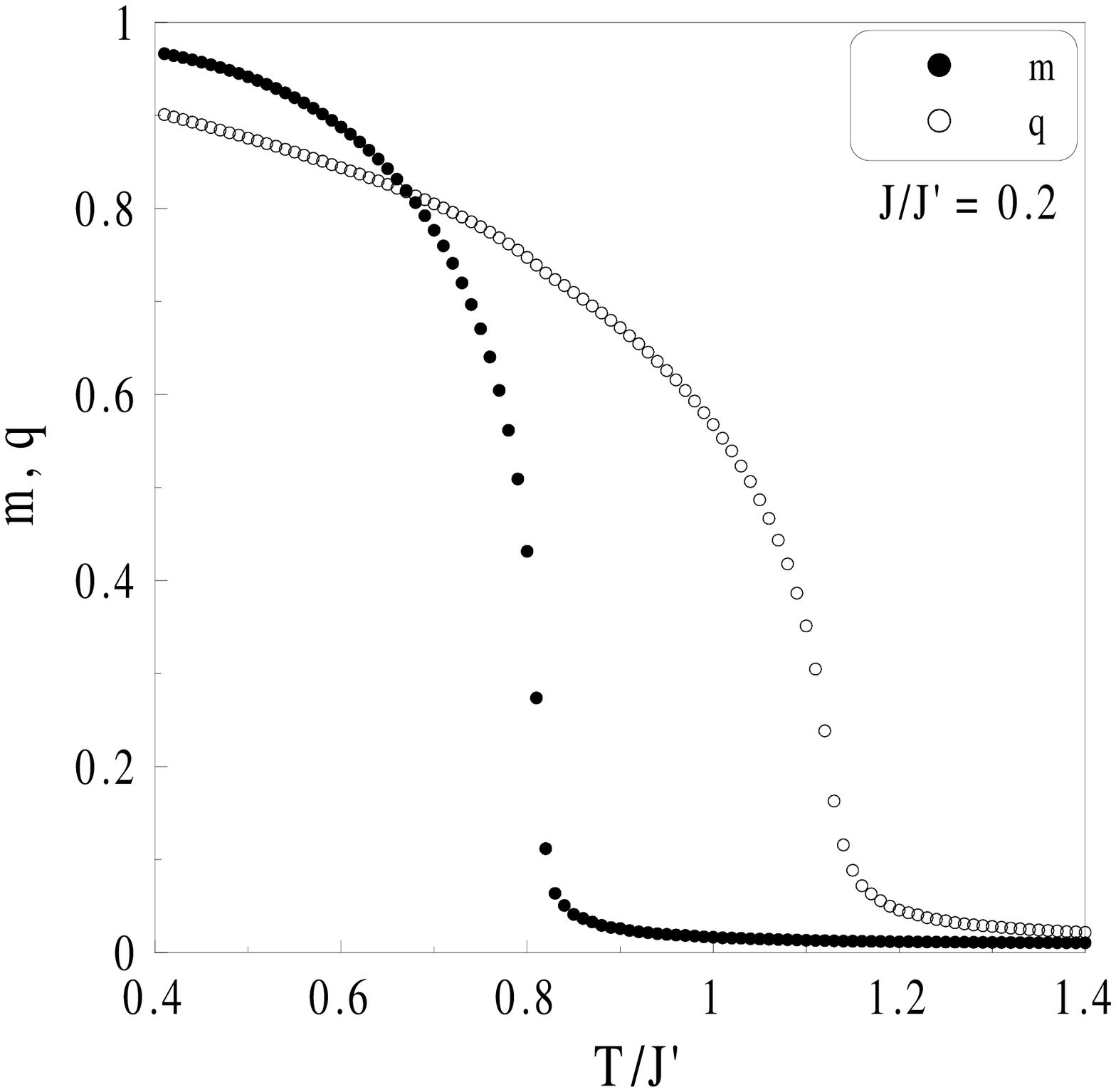}}
\subfigure{\includegraphics[scale=0.4]{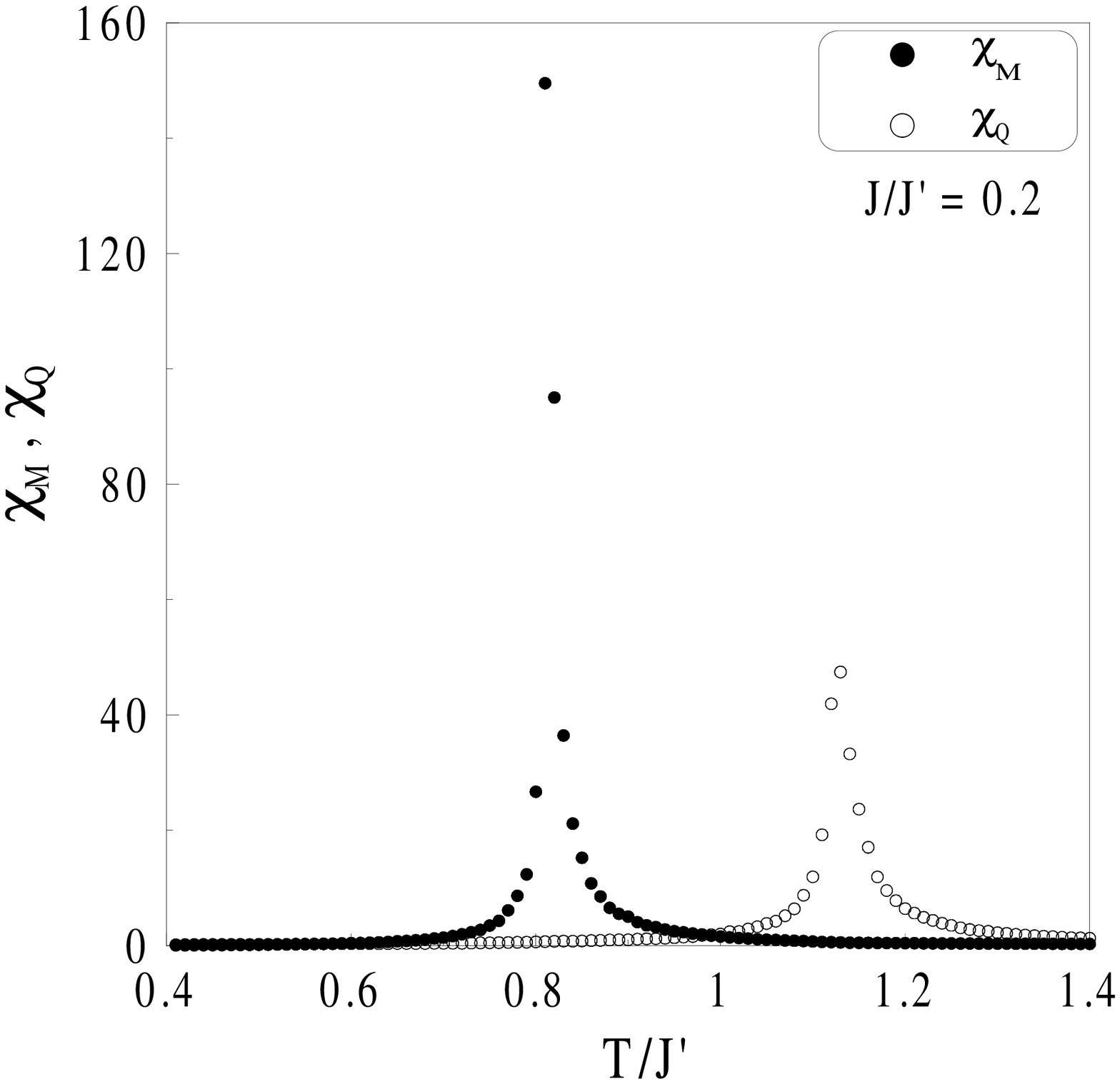}}
\caption{Temperature dependence of (a) the specific heat, (b) the DLRO and QLRO parameters
$m$ and $q$, respectively, and (c) the corresponding susceptibilities $\chi_{M}$ and $\chi_{Q}$ for
$J/J'$ = 0.2. Snapshots in Fig.(d) depict spin states in paramagnetic, QLRO and DLRO states at $T/J'$ =
1.4, 0.9 and 0.6, respectively.}
\end{figure}

\begin{figure}[!t]
\includegraphics[scale=0.5]{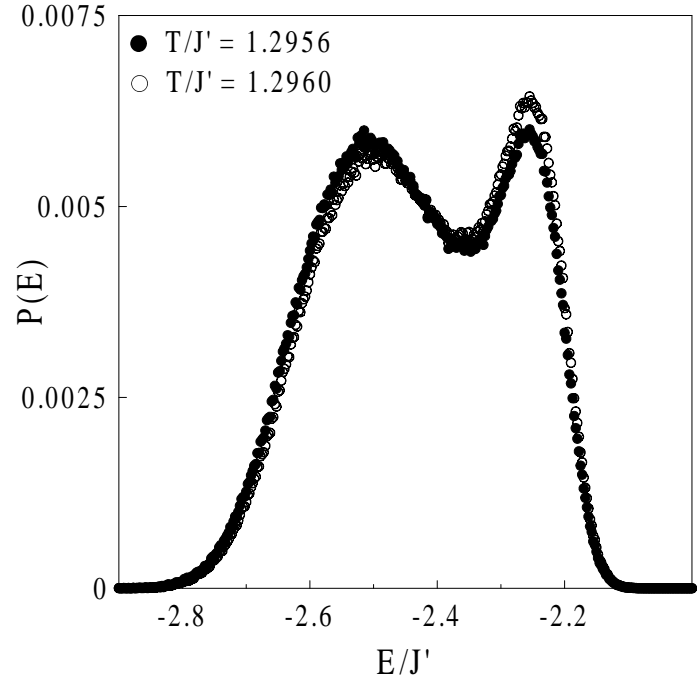}
\caption{Histogram of the internal energy distribution at the simulation temperature $T/J'$ =
1.2960 and after the peaks heights adjustment by lowering temperature to $T/J'$ = 1.2956.}
\end{figure}

\begin{figure}[!t]
\subfigure{\includegraphics[scale=0.4]{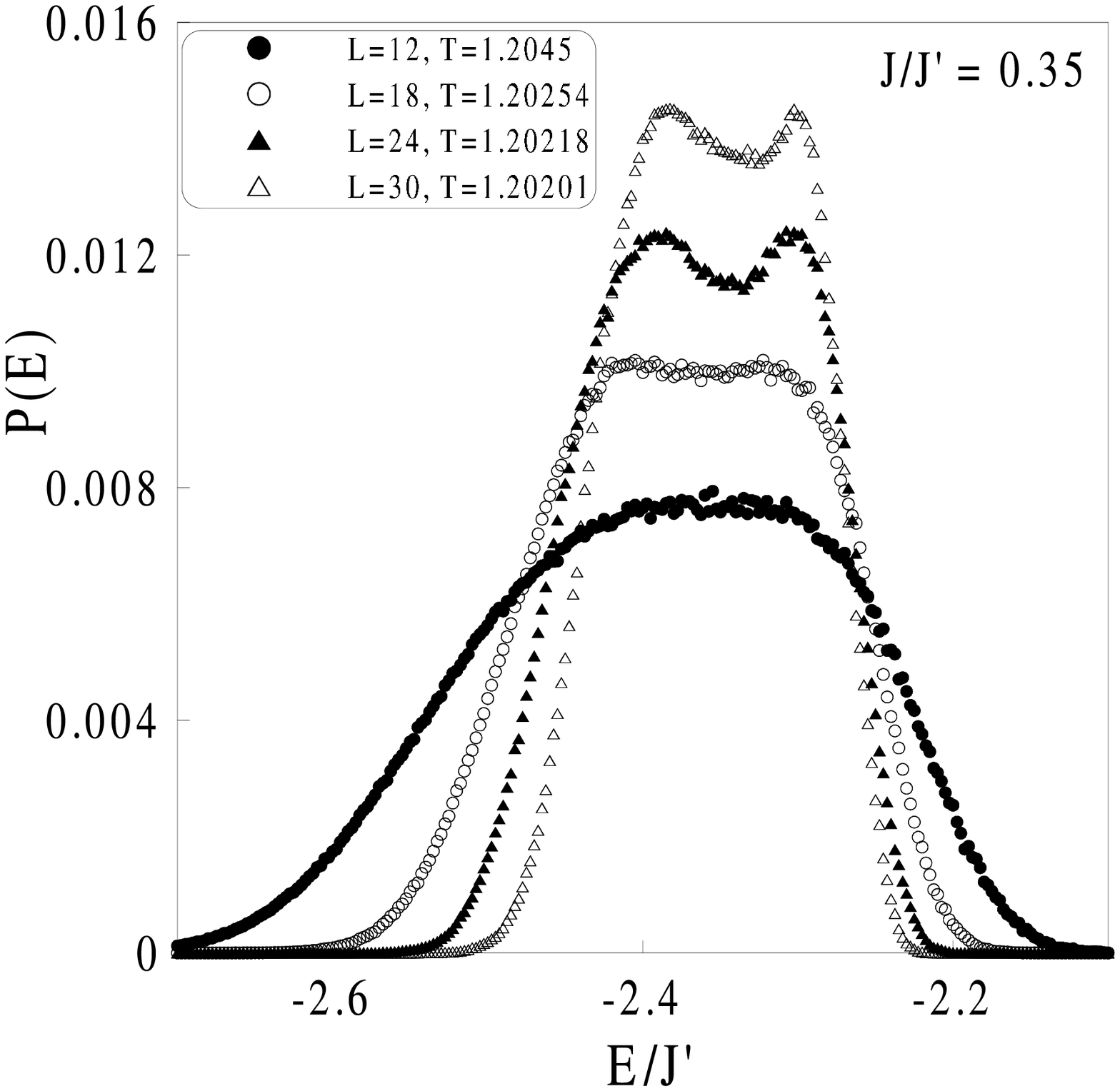}}
\subfigure{\includegraphics[scale=0.4]{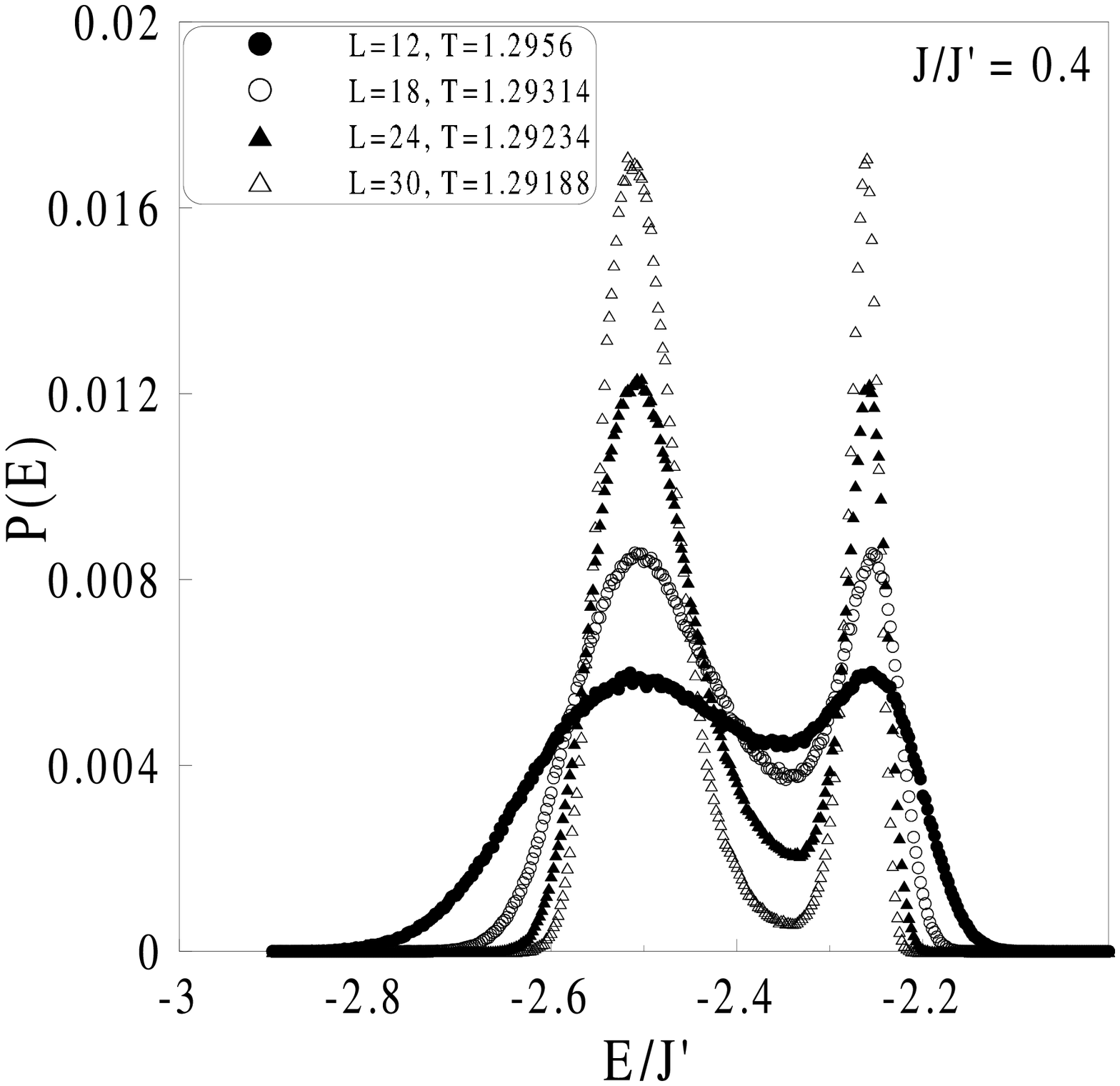}}
\subfigure{\includegraphics[scale=0.4]{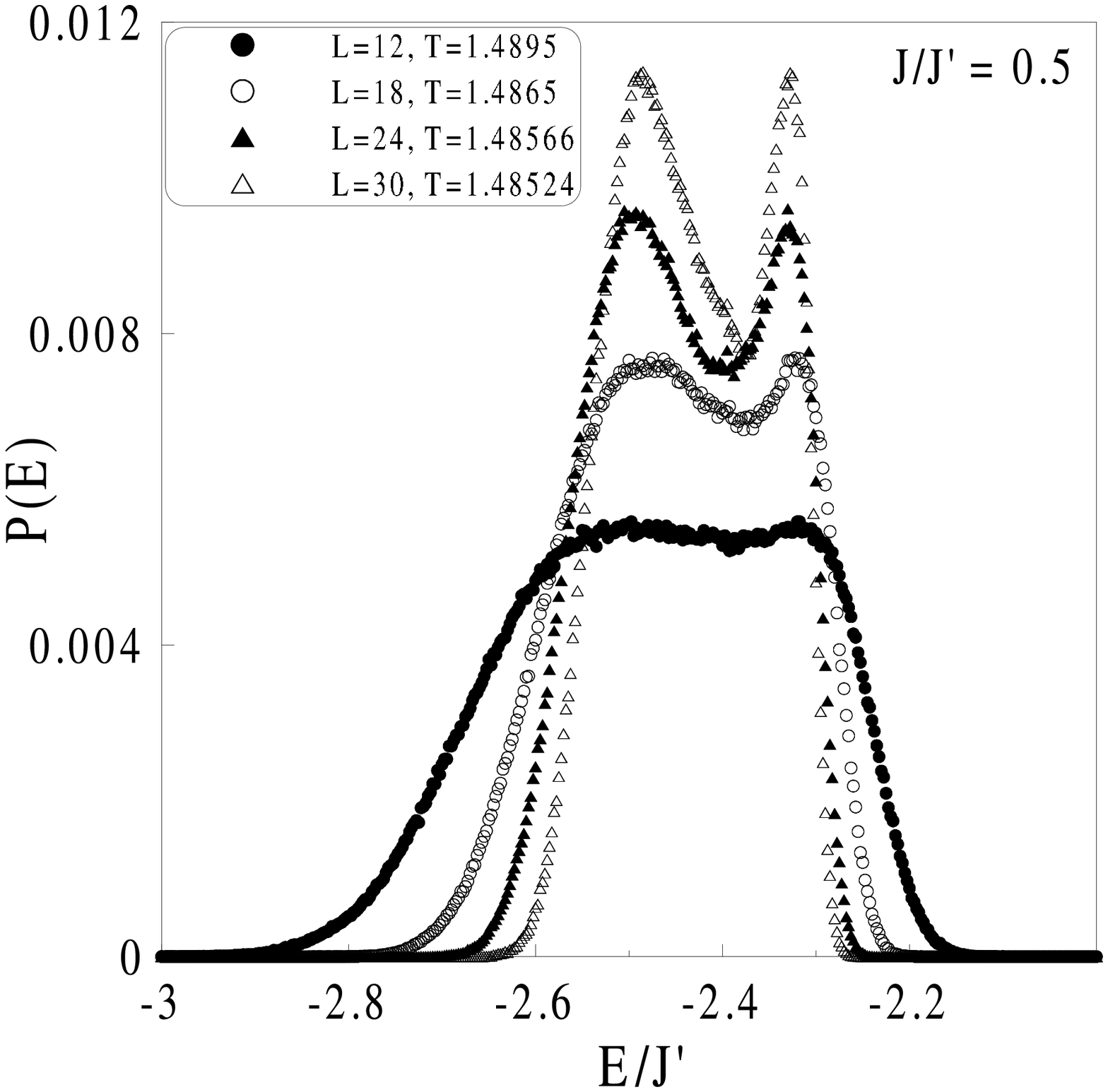}}
\caption{Energy distribution at the size-dependent transition temperatures $T_{c}(L)$ for
various lattice sizes and (a) $J/J'$ = 0.35, (b) 0.4, and (c) 0.5. Double-peak structures with deepening
barrier between the two energy states with increasing lattice size indicate first-order transitions.}
\end{figure}

\begin{figure}[!t]
\subfigure{\includegraphics[scale=0.4]{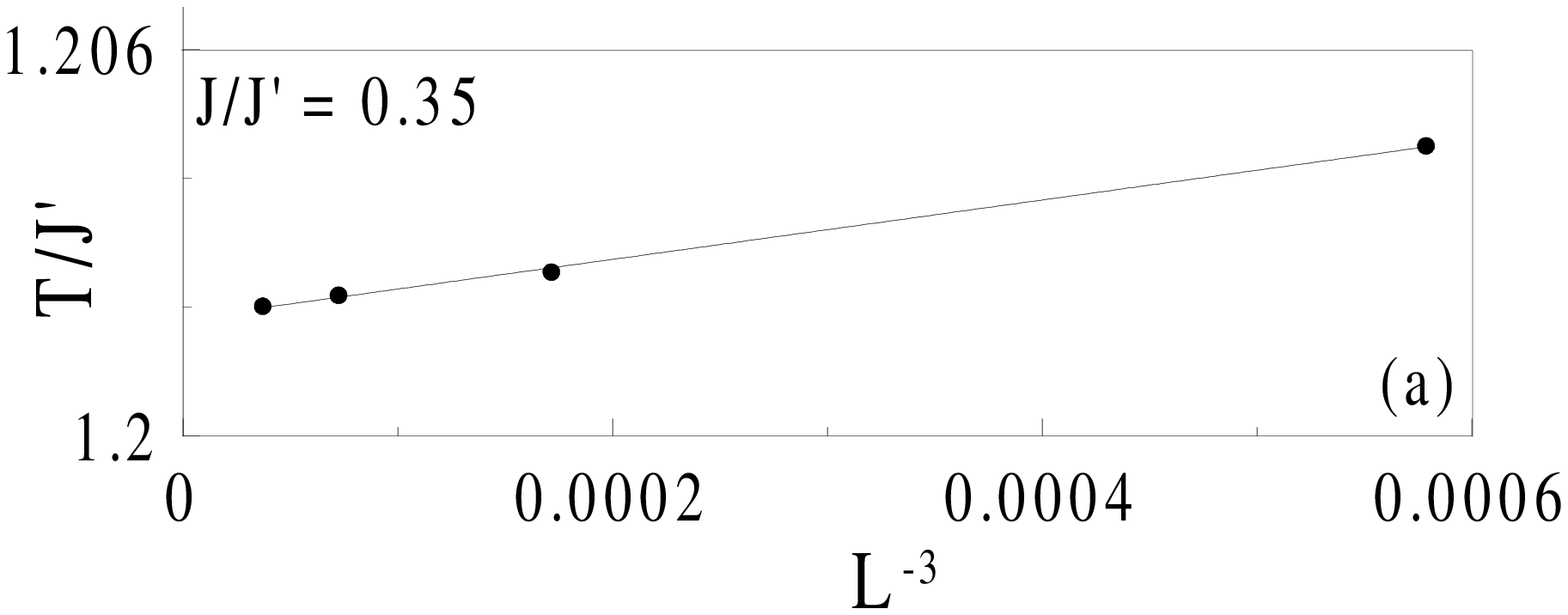}}
\subfigure{\includegraphics[scale=0.4]{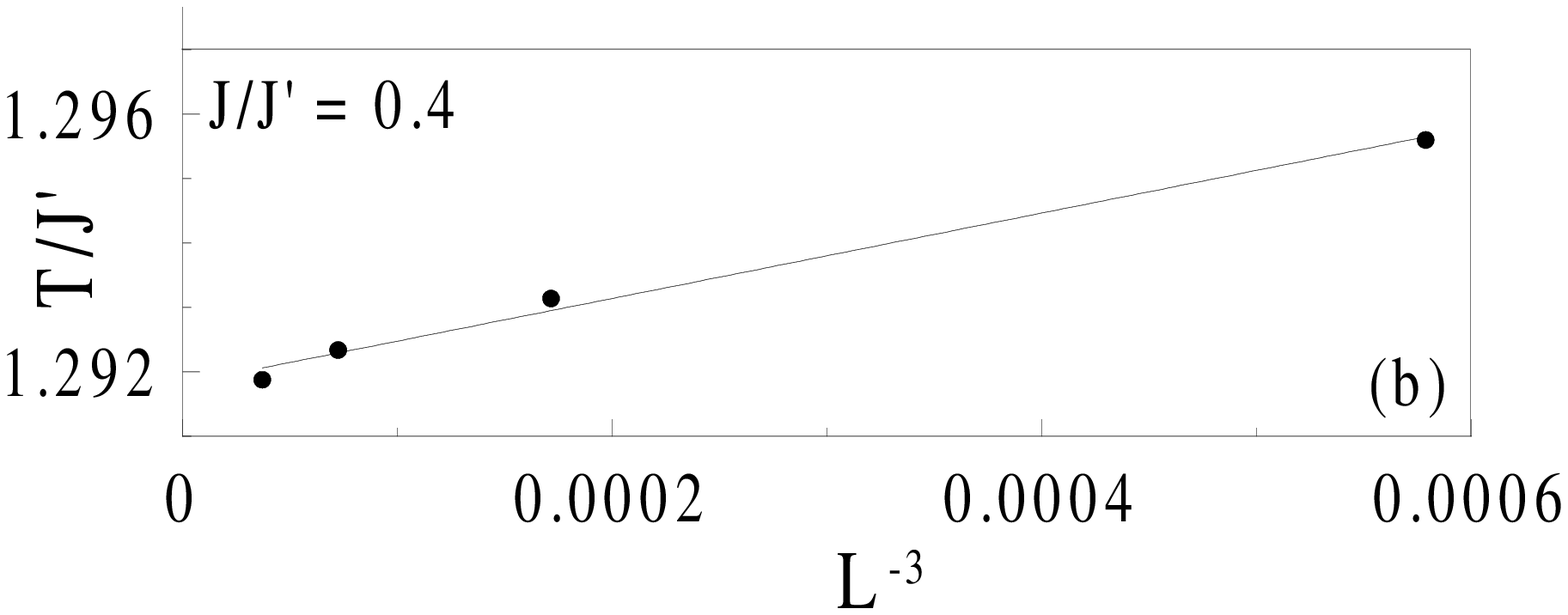}}
\subfigure{\includegraphics[scale=0.4]{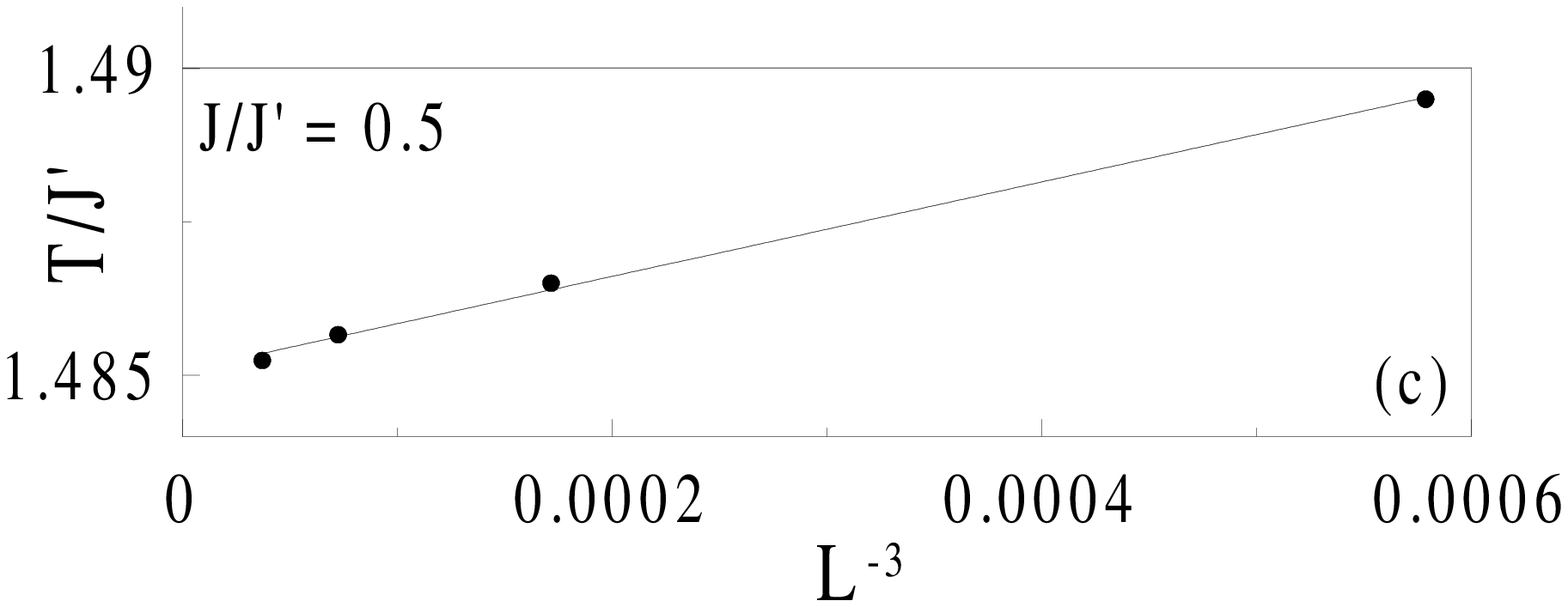}}
\caption{Volume dependence of the critical temperature for (a) $J/J'$ = 0.35, (b) 0.4, and (c) 0.5.}
\end{figure}

\begin{figure}[!t]
\includegraphics[scale=0.5]{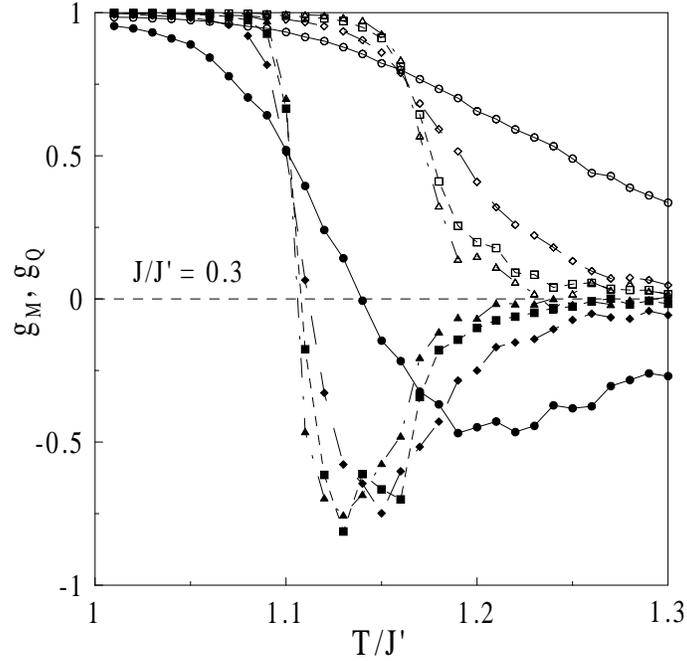}
\caption{Temperature variation of the Binder parameters $g_{O}(L,T)$ ($O$ = $M$, $Q$) for
$J/J'$ = 0.3. The circles, diamonds, squares and triangles represent data for the lattice sizes $L$ = 6,
12, 18 and 24, respectively, while the filled symbols represent $g_{M}$ and the blank ones $g_{Q}$ data.}
\end{figure}

\begin{figure}[!t]
\includegraphics[scale=0.5]{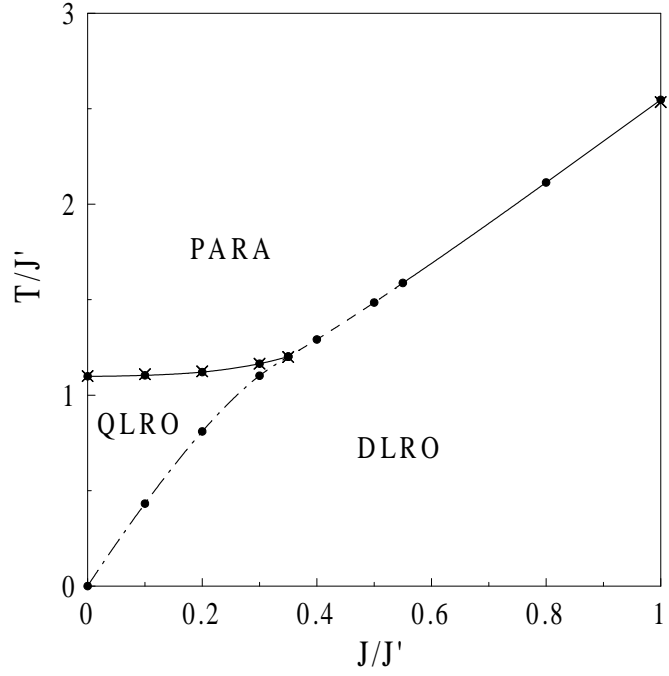}
\caption{Phase diagram in $(T/J',J/J')$ space. The solid and dashed lines denote second- and
first-order transitions, respectively. The order of the DLRO transition in  $0<J/J'<35$, denoted by the
dash-dot line, could not be determined with certainty. The crosses represent the HTSE results
\cite{chen-etal1,chen-etal2}}
\end{figure}

\begin{figure}[!t]
\includegraphics[scale=0.5]{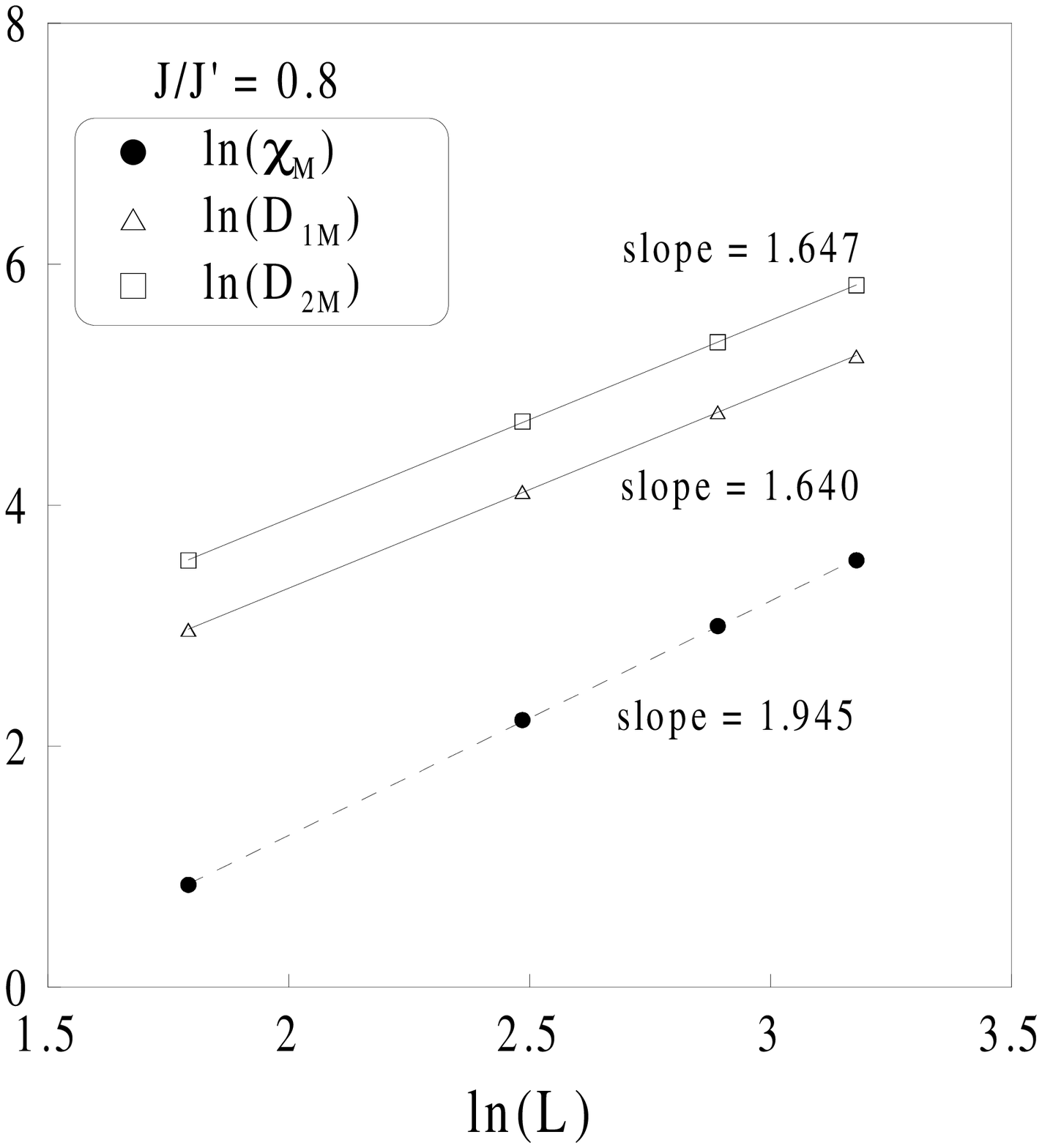}
\caption{Scaling behaviour of the maxima of the susceptibility $\chi_{M}$ and logarithmic
derivatives of the DLRO parameter and its second moment $D_{1M}$ and $D_{2M}$, respectively, in ln-ln
plot for $J/J'$ = 0.8. The slopes yield values of $1/\nu_{M}$ for $D_{1M},\ D_{2M}$ and
$\gamma_{M}/\nu_{M}$ for $\chi_{M}$.}
\end{figure}

\begin{figure}[!t]
\includegraphics[scale=0.5]{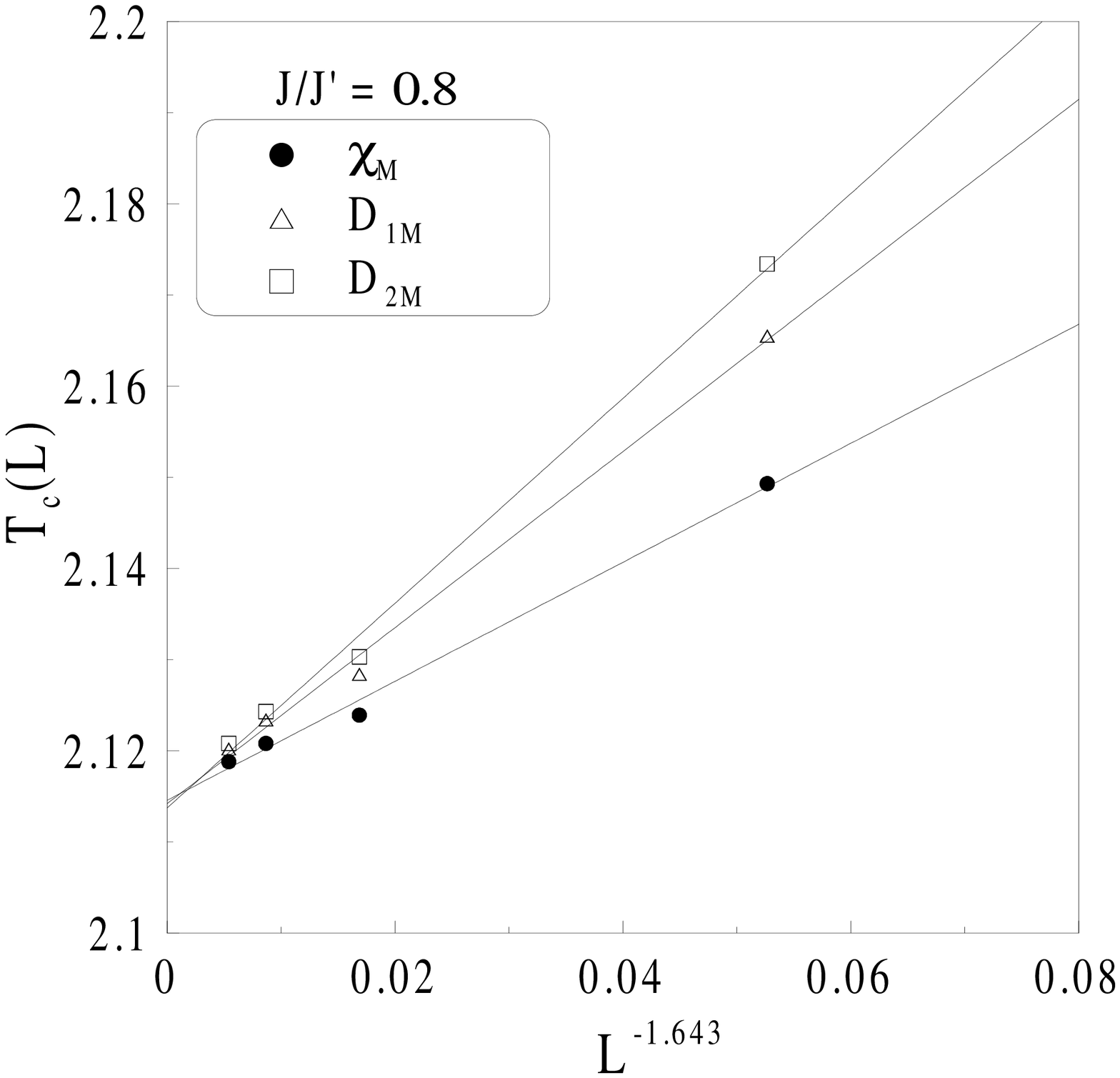}
\caption{Finite-size scaling of the critical temperature $T_{c}(L)$ determined by the peaks
of $\chi_{M}$, $D_{1M}$, and $D_{2M}$ for $J/J'$ = 0.8.}
\end{figure}

\begin{figure}[!t]
\includegraphics[scale=0.5]{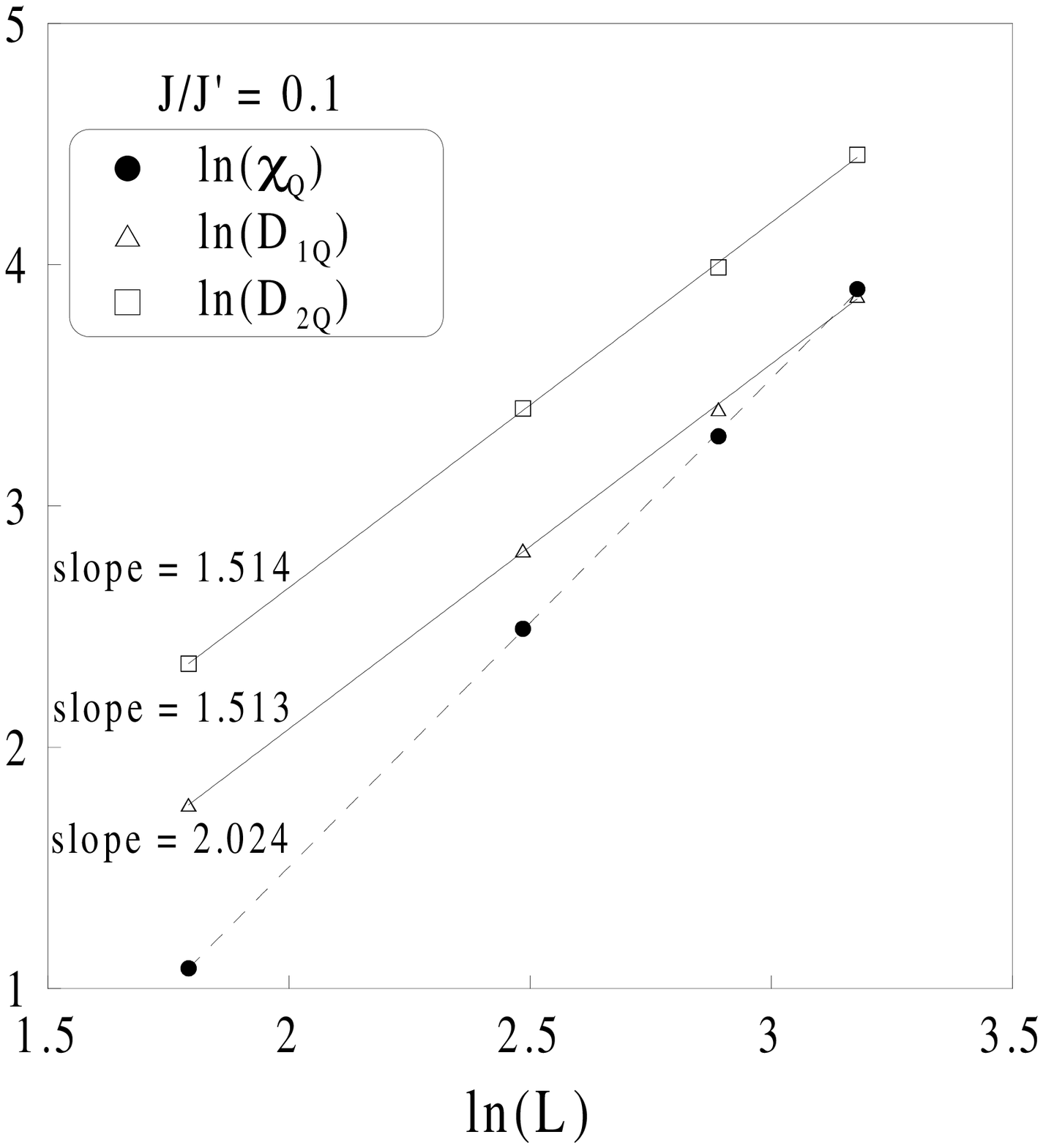}
\caption{Scaling behaviour of the maxima of the susceptibility $\chi_{Q}$ and logarithmic
derivatives of the QLRO parameter and its second moment $D_{1Q}$ and $D_{2Q}$, respectively, in ln-ln
plot for $J/J'$ = 0.1. The slopes yield values of $1/\nu_{Q}$ for $D_{1Q},\ D_{2Q}$ and
$\gamma_{Q}/\nu_{Q}$ for $\chi_{Q}$.}
\end{figure}

\begin{figure}[!t]
\includegraphics[scale=0.5]{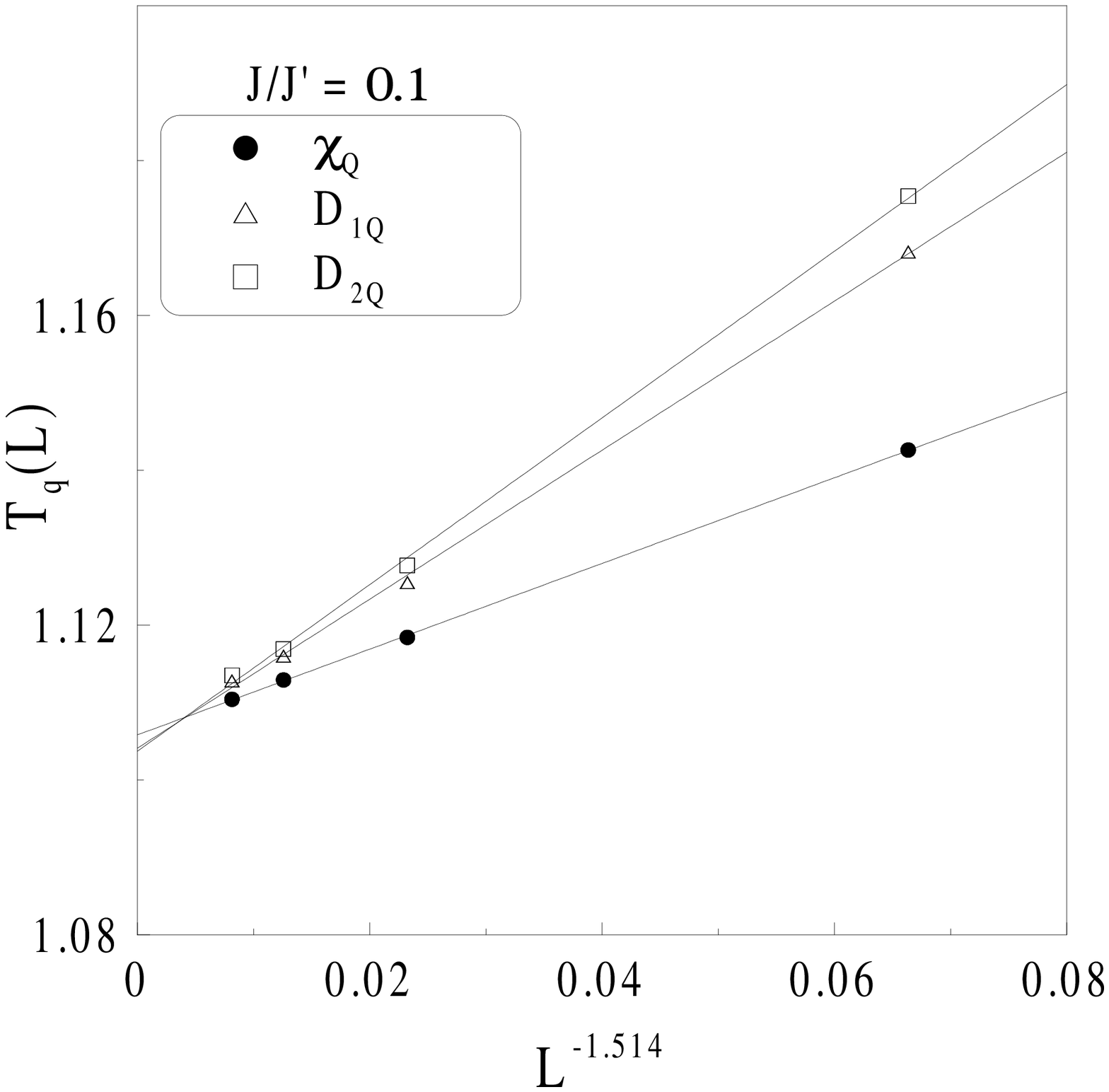}
\caption{Finite-size scaling of the critical temperature $T_{q}(L)$ determined by the peaks
of $\chi_{Q}$, $D_{1Q}$, and $D_{2Q}$ for $J/J'$ = 0.1.}
\end{figure}

\begin{figure}[!t]
\includegraphics[scale=0.5]{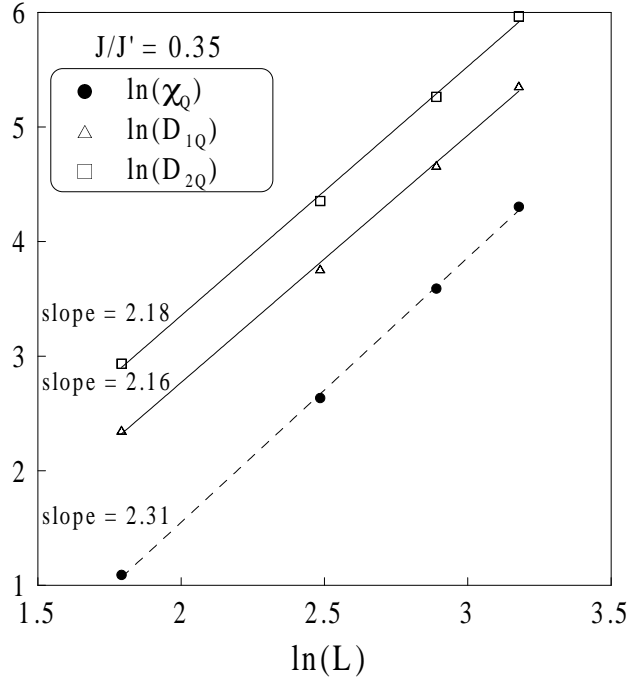}
\caption{Scaling behaviour of $\chi_{Q}$, $D_{1Q}$, $D_{2Q}$ for $J/J'$ = 0.35.}
\end{figure}

\begin{figure}[!t]
\includegraphics[scale=0.5]{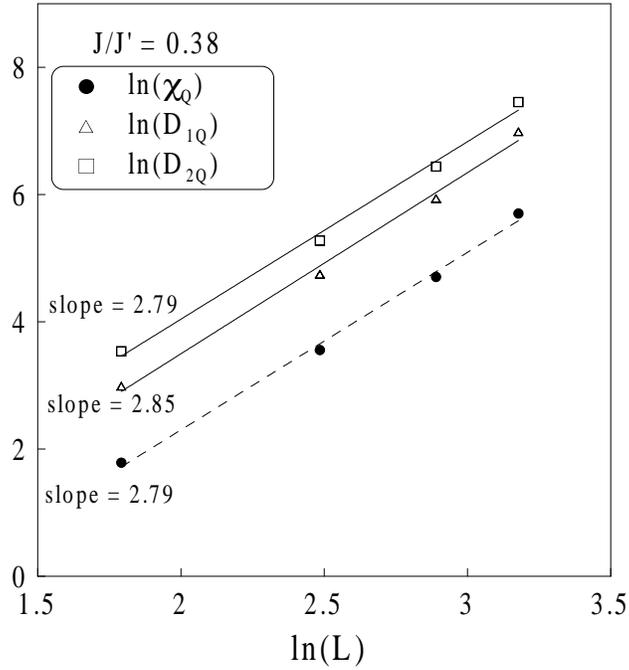}
\caption{Scaling behaviour of $\chi_{M}$, $D_{1M}$, $D_{2M}$ for $J/J'$ = 0.38. See text for
comments.}
\end{figure}


\vspace*{80mm}

\begin{table}[h]
\caption{The values of $T_c$, $\nu_{M}$ and $\gamma_{M}$.} \label{tab.1}
\begin{center}
\begin{tabular}{|c||c|c|c|c|c|}                                                                              \hline
 $J/J'$      & $T_{c}/J'$          & $\nu_{M}$       & $\gamma_{M}$  & $\gamma_{M}$ (HTSE)    \\ \hline\hline
 0.8         & 2.114(7)            & 0.608(11)       & 1.18(4)       &      -                 \\ \hline
 1.0         & 2.547(7)            & 0.630(31)       & 1.22(9)       & 1.14(2)        \\ \hline
 2.5         & 5.835(20)           & 0.649(9)        & 1.30(3)       & 1.25(2)        \\ \hline
 $\infty$    & 2.196(8)$^{*}$      & 0.669(38)       & 1.34(11)      & 1.33(2)        \\ \hline\hline
 $\infty$ Ref.\cite{janke}    & 2.202 $^{*}$        & 0.669(2)       & 1.32(1)       &                        \\ \hline
\end{tabular}
\newline
\end{center}
\hspace*{16mm}$^{*}$ Values of $T_{c}/J$ instead of $T_{c}/J'$
\end{table}


\vspace*{80mm}

\begin{table}[h]
\caption{The values of $T_q$, $\nu_{Q}$ and $\gamma_{Q}$.} \label{tab.2}
\begin{center}
\begin{tabular}{|c||c|c|c|c|c|c|}                                                                                 \hline
 $J/J'$      & $T_{q}/J'$          & $T_{q}/J'$ (HTSE) & $\nu_{Q}$       & $\gamma_{Q}$    & $\gamma_{Q}$ (HTSE)  \\ \hline\hline
 0           & 1.099(4)            & 1.100(5)          & 0.661(29)       & 1.34(7)         & 1.32(3)              \\ \hline
 0.1         & 1.104(4)            & 1.110(5)          & 0.661(38)       & 1.34(10)        & 1.32(3)              \\ \hline
 0.2         & 1.121(4)            & 1.125(5)          & 0.663(14)       & 1.34(6)         & 1.30(3)              \\ \hline
 0.3         & 1.165(3)            & 1.165(5)          & 0.649(42)       & 1.32(12)        & 1.26(3)              \\ \hline
\end{tabular}
\end{center}
\end{table}


\vspace*{80mm}

\begin{table}[h]
\caption{The slopes of $\chi_{O}$, $D_{1O}$ and $D_{2O}$, where $O = Q$ and $M$ for $J/J'=0.35$ and $0.38$, respectively.} \label{tab.3}
\begin{center}
\begin{tabular}{|c||c|c|c|}                                                                                     \hline
 $J/J'$      & slope of $\chi_{O}$         & slope of $D_{1O}$                & slope of $D_{2O}$    \\ \hline\hline
 0.35        & 2.31(10)                    & 2.16(15)                         & 2.18(15)                    \\ \hline
 0.38        & 2.79(32)                    & 2.84(40)                         & 2.79(36)                    \\ \hline
\end{tabular}
\end{center}
\end{table}

\end{document}